\documentclass[prl,twocolumn,preprintnumbers,nofootinbib,showpacs,amsmath,amssymb,superscriptaddress]{revtex4-1}

\usepackage{amsfonts,mathtools,amsthm}
\usepackage{amsmath}
\usepackage{amssymb}
\usepackage{upgreek}
\usepackage{bm}
\usepackage{dcolumn}
\usepackage{epsfig}
\usepackage{graphicx,xcolor}
\usepackage{graphics}
\usepackage[latin1]{inputenc}
\usepackage{latexsym}
\usepackage{rotating}
\usepackage[colorlinks=true, urlcolor=dblue, linktocpage=true, linkcolor=dblue,citecolor=purple]{hyperref}
\usepackage[all]{hypcap}
\usepackage{xspace}

\newcommand{\mrm}{\mathrm}

\newcommand{\nn}{\nonumber}

\newcommand{\lb}{\left(}
\newcommand{\rb}{\right)}

\newcommand{\OO}{\mathcal{O}}

\newcommand{\lsb}{\left[}
\newcommand{\rsb}{\right]}

\definecolor{dgreen}{RGB}{17,96,98}

\newcommand\Qp{{\mathbb{Q}_p}}
\newcommand\Zp{{\mathbb{Z}_p}}

\colorlet{dblue}{blue!70!black}

\newcommand{\arxivold}[1]
  {\href{http://arxiv.org/abs/#1}{#1}}
\newcommand{\arxiv}[1]
  {\href{http://arxiv.org/abs/#1}{arXiv:#1}}

\begin{document}

\title{Closed form expression for the 5-point Veneziano amplitude in terms of 4-point amplitudes}

\author{Bogdan Stoica}
\affiliation{Department of Physics \& Astronomy, Northwestern University, Evanston, Illinois 60208, USA}

\date{\today}

\begin{abstract} 

\noindent We propose a number system covariance principle between $p$-adic and Archimedean frameworks. We use it to derive several closed-form expressions for the five-point open string tachyon scattering~amplitude.
\end{abstract}


\preprint{nuhep-th/21-11}

\maketitle

\noindent In this Letter we will propose a novel organizational principle for physical prediction,  namely that of number system covariance between the $p$-adic and real numbers. This will allow us to recurse the predictions of higher-derivative corrections, which we apply to the  5-point tachyonic open string tree level amplitude. This principle suggests that the amplitudes (and other physical quantities) should obey the same form for theories defined on $\mathbb{R}$ and on the $p$-adic fields $\Qp$, and furthermore should obey the same relations, when cast in appropriate form. This principle is in concordance with general relativity, where covariant expressions take the same form in any coordinate system expressed in the same mathematical framework. Here we generalize to a sense of universality between the $p$-adic and Archimedean frameworks.  

Number system covariance is already exhibited by many physical quantities, such as free propagators in quantum mechanics and field theory \cite{vvzbook,Huang:2020aao,Stoica:2018zmi}, and $4$-point open string tachyon amplitudes \cite{FreundWitten}: The physical expressions take the same form, under appropriate substitutions, such as the Archimedean absolute value norm going to $p$-adic norm, or the usual Fourier factor being replaced by the $p$-adic Fourier factor. In fact, in these examples an even stronger relation holds, that of a unit Euler product across all places. However, the Euler product relation no longer holds for the 5-point Veneziano amplitudes \cite{BFOW}.  Rather more intricate relations, that hold only in special situations, have been identified \cite{adelicNpoint}.

In the present work we will use number system covariance to derive novel recursive relations between 5-point and 4-point amplitudes that hold in both $p$-adic and Archimedean frameworks, and yield a closed form description for the 5-point Archimedean Veneziano amplitudes. These relations can be expanded order by order to arbitrarily higher derivative corrections in $\alpha'$. Since all $p$-adic $n$-point Veneziano amplitudes can be computed in closed form, this may suggest a path towards expressing $n$-point  Archimedean Veneziano amplitudes in terms of products of lower point amplitudes and special functions.\footnote{For $p$-adic string models see~\cite{zabrodin,zabrodin2,Huang:2019nog}.}

\emph{Summary of results.}--- We will find that in both the $p$-adic and Archimedean cases the 5-point Veneziano amplitude takes the form
\begin{equation}
\label{eq1}
A_5(k_j) = \sum_i A_4\lb S^{(1)}_{i},T^{(1)}_{i}\rb A_4\lb S^{(2)}_{i},T^{(2)}_{i}\rb Z_i(\{k_j\}_i),
\end{equation}
where $A_5(k_j)$ is the 5-point Veneziano open string tachyon scattering amplitude summed over all momentum permutations, and  $A_4$ are the 4-point Veneziano amplitudes
\begin{equation}
\label{eq2}
A_4(S,T)= \int_\mathbb{K} |x|^S|1-x|^T dx = \frac{\Gamma(S+1)\Gamma(T+1)}{\Gamma(S+T+2)},
\end{equation}
applied to certain combinations of the kinematic parameters. For the usual (Archimedean) Veneziano amplitudes $|\cdot|$ is the absolute value norm and $\mathbb{K}=\mathbb{R}$; for the $p$-adic open string amplitudes $\mathbb{K}=\Qp$ and $|\cdot|$ is the $p$-adic norm. The \emph{dressing functions} functions $Z$ are simple functions of one argument on the $p$-adic side, and $_3F_2$ or $_6F_5$ hypergeometric functions of kinematic parameters, at the critical point $z=1$, on the Archimedean side. In fact, on the Archimedean side it is possible to rearrange Eq. \eqref{eq1} in several ways, changing the hypergeometric functions and their arguments, as well as the integration domain from $\mathbb{R}$ to $(0,1)$ and other intervals, which we will explain~below.

The Gamma functions on the right-hand side of Eq.~\eqref{eq2} correspond to the integration domain: i) if $\mathbb{K}=\Qp$ then $\Gamma$ is the $p$-adic Gelfand-Graev Gamma function (and $A_4$ is the $p$-adic open string amplitude), ii) if $\mathbb{K}=(0,1)$ (corresponding to $A_4$ being the contribution of one ordering of tachyon vertex operator insertions on the open string worldsheet boundary) it is the Euler Gamma function, and iii) if $\mathbb{K}=\mathbb{R}$ (corresponding to $A_4$ being the permutation-invariant tachyon scattering amplitude summed over all vertex operator insertion orderings) then $\Gamma$ is the Archimedean Gelfand-Graev Gamma function corresponding to the absolute value $|\cdot|^s$, related to the Euler Gamma function by normalization~as
\begin{equation}
\Gamma_{G}(s) = 2^{1-s} \pi ^{-s} \cos \left(\frac{\pi  s}{2}\right) \Gamma_E(s).
\end{equation}
\emph{Notation:} We will use superscript (or subscript)~$p$ to denote the $p$-adic quantities, $E$ to denote the Archimedean Euler Gamma quantities (corresponding to the Veneziano amplitude associated to one ordering of the vertex operator insertions on the worldsheet boundary), and $G$ to denote the Archimedean Gelfand-Graev quantities (corresponding to the Veneziano amplitude summed over all momenta permutations). Note that in the $p$-adic case there are no ordered contributions to the amplitude, because the boundary of the $p$-adic worldsheet (the Bruhat-Tits tree) is unordered.

The number system covariance between the $p$-adic and Archimedean sides also holds for the derivation of Eq.~\eqref{eq1}, since on both the $p$-adic and Archimedean sides this equation is obtained by applying the same Fourier trick involving multiplicative characters, and all manipulations are precisely parallel.

The $p$-adic 5-point tachyon open string amplitude, in Koba-Nielsen form, is given by \cite{Virasoro1968,BardakciRuegg,KobaNielsen1969,FairlieJones}
\begin{eqnarray}
\label{eq3}
A^{(p)}_5(k_i) &=& \int_{\mathbb{Q}_p^2} |x_3|_p^{S_2} |x_3|_p^{S_3}|1-x_2|_p^{T_2}|1-x_3|_p^{T_3}\times\\
& &\times|x_2-x_3|_p^{M_{23}}dx_2dx_3. \nn
\end{eqnarray}
Here index $i$ runs over the five external particles, conservation of momentum dictates $\sum_{i=1}^5 k_i =0$, and for kinematic variables we have chosen the convention
\begin{eqnarray}
S_{2,3} &\coloneqq& \alpha' k_1\cdot k_{2,3}, \nn \\
T_{2,3} &\coloneqq& \alpha' k_4\cdot k_{2,3}, \\
M_{23} &\coloneqq& \alpha' k_2\cdot k_3. \nn
\end{eqnarray}

Because the field $\Qp$ can be partitioned into disjoint subsets of constant norm, the integrals in Eq. \eqref{eq3} can be evaluated directly, leading to a well-known set of Feynman rules for the $p$-adic amplitudes, as well as a resummation of the amplitudes into an effective action \cite{BFOW}. Thus, Eq.~\eqref{eq1} could in principle be checked by direct computation. However, this approach is not number system covariant, because the same integrals cannot be directly performed on the Archimedean side. In order to ensure an evaluation that applies for both $\mathbb{R}$ and $\Qp$, we must employ a Fourier trick for integrating norms. For $\Qp$, this Fourier trick takes the~form
\begin{eqnarray}
\label{FTtrickQp}
& &\int_\Qp |x|_{p}^S |1-x|_{p}^T e^{2\pi i \{ k x\}} dx =\\
&=&\begin{cases}
A^{(p)}_4\lb S,T \rb + \frac{\Gamma_{p}\lb S+T+1 \rb}{|k|_p^{S+T+1}} \qquad\quad\, \mrm{if\ } |k|_p\leq 1\\
\frac{\Gamma_{p}\lb S+1 \rb}{|k|_p^{S+1}} + \frac{\Gamma_p\lb T+1 \rb}{|k|_p^{T+1}} e^{2\pi i \{k\}_p} \qquad\, \mrm{if\ } |k|_p>1
\end{cases}.\nn
\end{eqnarray}
Here $\Gamma_{p}(s)$ is the $p$-adic Gelfand-Graev Gamma function
$\Gamma_{p}(s) = (1-p^{s-1})/(1-p^{-s}),$
and $A^{(p)}_4(S,T)$ is given by Eq. \eqref{eq2} in the $p$-adic case $\mathbb{K}=\Qp$, i.e. it is the permutation symmetric $p$-adic tachyon scattering amplitude. The function $e^{2\pi i \{kx\}_p}$ is the $p$-adic Fourier factor, with $\{\cdot\}_p$ the $p$-adic factional part (for a review see e.g.~\cite{GGIP}). Eq.~\eqref{FTtrickQp} follows from straightforward computation of the integrals, by partitioning $\Qp$ into subsets of constant norm (see \cite{paper2} for details).

The Fourier trick \eqref{FTtrickQp} can be used to evaluate the integrals in \eqref{eq3}, by first replacing the norm $|x_2-x_3|_p^{M_{23}}$ by its Fourier transform,
\begin{eqnarray}
A^{(p)}_5(k_i) &=& \frac{1}{\Gamma_p\lb-M_{23}\rb} \int_{\mathbb{Q}_p^2}  |x_2|_p^{S_2} |x_3|_p^{S_3} |1-x_2|_p^{T_2} |1-x_3|_p^{T_3} \times\nn\\
& & \times |k|_p^{-M_{23}-1} e^{2\pi i \{ k\lb x_3-x_2 \rb \}_p} dkdx_2dx_3,
\end{eqnarray}
and then using Eq. \eqref{FTtrickQp} to compute the $x_{2,3}$ integrals. Finally, the $k$ integral can be performed, yielding (let $G_2\coloneqq S_2 + T_2+M_{23} +1$, $G_3\coloneqq S_3 + T_3+M_{23}+1$)
\begin{widetext}
\begin{eqnarray}
&&A^{p}_5(k_i) = A^{p}_4(S_2,T_2) A^p_4(S_3,T_3) Z_p\lb M_{23} \rb+ A^p_4\lb S_2 + T_2 , S_3 + T_3 \rb A^p_4\lb S_2 +T_2 + S_3 + T_3 +1 , M_{23} \rb Z_p \lb G_2 + G_3 - M_{23} \rb \nn\\
&+& A^p_4\lb S_3,T_3 \rb A^p_4\lb  S_2+T_2, M_{23} \rb Z_p\lb G_2\rb  + A^p_4\lb S_2,T_2 \rb A^p_4\lb  S_3+T_3, M_{23} \rb Z_p\lb G_3\rb  \nn\\
&-&  A^p_4\lb S_2, S_3 \rb A^p_4\lb S_2 + S_3+1, M_{23} \rb Z_p\lb S_2 + S_3 +M_{23} + 2 \rb -  A^p_4\lb T_2, T_3 \rb A^p_4\lb T_2 + T_3+1, M_{23} \rb Z_p\lb T_2 + T_3 +M_{23} + 2 \rb \nn\\
&+&A^p_4\lb S_2, T_3  \rb A^p_4 \lb S_2+T_3 + 1, M_{23} \rb \lb 1 - Z_p\lb S_2 + T_3 +M_{23} + 2 \rb \rb\nn\\
\label{eq8llong}
&+&A^p_4\lb S_3, T_2  \rb A^p_4 \lb S_3+T_2 + 1, M_{23} \rb \lb 1 - Z_p\lb S_3 + T_2 +M_{23} + 2 \rb \rb,
\end{eqnarray}
\end{widetext}
where the dressing function $Z_p$ is given by (note the appearance of the local zeta factor $\zeta_p(s)=1/(1-p^{-s})$)
\begin{equation}
Z_p(s) = \frac{p-1}{p-p^{-s}}.
\end{equation}
Equation \eqref{eq8llong} is the $p$-adic 5-point open string tachyon amplitude, as a sum of double-copies of 4-point amplitudes. Of course, this answer matches the result that can be derived from the Feynman rules in \cite{BFOW}, however writing it recursively will make manifest the number system covariance with the Archimedean framework.\footnote{For analytic properties of the Koba-Nielsen integrals see \cite{Bocardo-Gaspar:2016zwx,Bocardo-Gaspar:2017atv,Bocardo-Gaspar:2019pzk}.}

The appearance of the recursive structure in Eq.~\eqref{eq8llong} can be understood mathematically as follows. The Fourier trick contains a factor of $A^{(p)}_4(S,T)$, coming from the integral of $|x|_p^S|1-x|_p^T$ when the Fourier factor is $1$ i.e. on the set of $p$-adic integers $\Zp$ inside $\Qp$. Then the two factors of $|x|_p^S|1-x|_p^T$ in $A^{(p)}_5(k_i)$ require that the Fourier trick be applied twice, naturally giving rise to the double-copy of four-point amplitudes.

As explained above, the treatment of the Archimedean computation is perfectly parallel to the $p$-adic one, manifesting our proposed number system covariance. Thus, the first step to obtaining a closed form expression for $A_5(k_i)$ on the Archimedean side is to establish the Fourier trick. This takes the form
\begin{widetext}
\begin{eqnarray}
\label{exp10}
& &\int_{-\infty}^\infty |x|^S|1-x|^T e^{-2\pi i k x} dx = A^{(G)}_4(S,T)\, {}_2F_3\lb  \frac{S+1}{2} , 1 + \frac{S}{2} ; \frac{1}{2}, 1 + \frac{S+T}{2}, \frac{3+S+T}{2} ; -\pi^2 k^2 \rb \\
& & + 2 i \pi k  \tan \frac{\pi S}{2} \tan \lb \frac{\pi(S+T)}{2} \rb A^{(G)}_4(S+1,T) \, {}_2F_3\lb 1 + \frac{S}{2}, \frac{3+S}{2} ;  \frac{3}{2}, \frac{3+S+T}{2}, 2 + \frac{S+T}{2}; -\pi^2k^2 \rb \nn\\
& & + \frac{k}{|k|^{S+T+1}} \frac{iT \tan \frac{\pi\lb S+T \rb}{2} }{\Gamma_G\lb 1-S-T \rb} \, {}_2F_3\lb \frac{1-T}{2}, 1 - \frac{T}{2} ; \frac{3}{2} , \frac{1-S-T}{2}, 1 - \frac{S+T}{2} ; -\pi^2 k^2 \rb \nn\\
& & + \frac{1}{|k|^{S+T+1}\Gamma_G\lb -S-T \rb} \, {}_2F_3\lb \frac{1-T}{2}, - \frac{T}{2} ; \frac{1}{2} , -\frac{S+T}{2}, \frac{1-S-T}{2} ; - \pi^2k^2 \rb. \nn
\end{eqnarray}
\end{widetext}
Expression \eqref{exp10} is the Archimedean analogue of the $p$-adic formula \eqref{FTtrickQp}: note in both cases the appearance of the $A_4$ amplitude. The integral in Eq. \eqref{exp10} converges in the region $S,T>-1$, $S+T<-1$, and can be analytically continued outside this region (for a table of integrals see e.g. \cite{GradRyzhik}).

In the Archimedean case, the Koba-Nielsen integral expression  for the full amplitude is the same as Eq. \eqref{eq3}, but with the $p$-adic norms replaced by absolute value norms, and $\Qp$ replaced by $\mathbb{R}$. Writing the norm $|x_2-x_3|^{M_{23}}$ in terms of its Fourier transform (analogously to Eq. \eqref{eq3} in the $p$-adic case) gives
\begin{eqnarray}
A^{(G)}_5(k_i) &=& \frac{1}{\Gamma_G\lb-M_{23}\rb} \int_{\mathbb{R}^2} |x_2|^{S_2} |x_3|^{S_3} |1-x_2|^{T_2}  \times\\
& & \times |1-x_3|^{T_3} |k|^{-M_{23}-1} e^{2\pi i  k\lb x_3-x_2 \rb} dkdx_2dx_3. \nn
\end{eqnarray}
Just as in the $p$-adic case, the $x_2$ and $x_3$ integrals can then be evaluated by the Fourier trick \eqref{exp10}, and then the $k$ integral can be performed, yielding a double-copy relation of the form \eqref{eq1}, with the sum over $10$ terms and the $Z_i$ dressing functions equal to hypergeometric $_6F_5$ functions multiplied by combinations of trigonometric functions, possibly times a linear term (the precise expression is given in the auxiliary Mathematica file). This expression is an Archimedean analog of Eq. \eqref{eq8llong}. It is fully permutation symmetric in the momenta, though this is not readily apparent from the explicit expression.

Let's now comment on the relation between the $p$-adic and Archimedean closed-form expressions for the amplitude. Although the steps in the derivations and general form of the expressions are the same on the $p$-adic and Archimedean side, the $Z$ dressing functions are very different on the two sides. This is a feature, not a bug, because Archimedean expressions can be considerably more involved than $p$-adic ones, and part of the usefulness of the number system covariance principle is that it can account for this.

It is possible to use the strategy above to further obtain alternative (and simpler) closed form expressions for the $A_5(k_i)$ amplitude. To do this, one can split the integration domain $\mathbb{R}^2$ as
\begin{equation}
\label{eqR2}
\mathbb{R}^2 = ((-\infty,0)\cup (0,1)\cup(1,\infty))^2,
\end{equation} 
and integrate on the subdomains separately. We have an analogue of the Fourier trick as
\begin{eqnarray}
\label{FTtrick11}
& &\int_0^1 |x|^S|1-x|^T e^{-2\pi i k x} dx = \\
& &=A^{(E)}(S,T) {}_1F_1(S+1,S+T+2,-2\pi i k), \nn
\end{eqnarray}
and similarly
\begin{eqnarray}
\label{FTtrick22}
& &\int_1^\infty |x|^S|1-x|^T e^{-2\pi i k x} dx = \frac{\Gamma_E (S+T+1)}{(2\pi ik)^{S+T+1}} \times\\
& &\times _1F_1(-T;-S-T;-2 i k \pi )-\frac{\sin \pi  S}{\sin\pi(S+T)} A_4^{(E)}(S,T)\times \nn\\
& &\times _1F_1(S+1;S+T+2;-2 i k \pi ).\nn
\end{eqnarray}
The integrals on all other subdomains can be related by coordinate transformations to expressions \eqref{FTtrick11}, \eqref{FTtrick22}, and the $k$ integrals can be performed. This yields an alternate expression for the 5-point amplitude, this time in terms of $_3F_2$ hypergeometric functions,
\begin{widetext}
\begin{eqnarray}
A_5^{(G)}(k_i) &=& A^{E}_4(S_2,T_2) \lsb A^E_4(T_3,M_{23}+S_3)+A^E_4(T_3,-G_3-1)\rsb \, _3F_2(-M_{23},S_2+1,-G_3;-M_{23}-S_3,S_2+T_2+2;1)\nn\\
&+&A^E_4(-G_2-1,S_2) A^E_4(T_3,-G_3-1) \, _3F_2(-M_{23},S_2+1,T_3+1;-M_{23}-S_3,-M_{23}-T_2;1)\nn\\
&+&\lsb A^E_4(S_3,T_3) A^E_4(T_2,-G_2-1)+A^E_4(T_3,-S_3-T_3-2) A^E_4(-G_2-1,T_2)\rsb \times\nn\\
& &\times _3F_2(-M_{23},S_3+1,-G_2;-M_{23}-S_2,S_3+T_3+2;1)\nn\\
&+&A^E_4(T_2,-G_2-1) A^E_4(-G_3-1,S_3) \, _3F_2(-M_{23},S_3+1,T_2+1;-M_{23}-S_2,-M_{23}-T_3;1)\nn\\
&+&\lsb A^E_4(S_3,-S_3-T_3-2) A^E_4(-G_2-1,S_2)+A^E_4(T_3,S_3) A^E_4(S_2,-G_2-1)\rsb\times\nn\\
& &\times _3F_2(-M_{23},-G_2,T_3+1;-M_{23}-T_2,S_3+T_3+2;1)\nn\\
&+&A^E_4(T_2,S_2) A^E_4(S_3,-G_3-1) \,_3F_2(-M_{23},T_2+1,-G_3;S_2+T_2+2,-M_{23}-T_3;1)\nn\\
&+& \lb 1- \frac{\sin \pi S_3}{ \sin \pi (M_{23}+S_3)}\rb A^E_4(S_3,M_{23}) A^E_4(T_2,M_{23}+S_2+S_3+1) \times\nn\\
& &\times _3F_2(S_3+1,M_{23}+S_2+S_3+2,-T_3;M_{23}+S_3+2,G_2+S_3+2;1)\nn\\
&+& \lb 1- \frac{\sin \pi (M_{23}+S_3+T_3)}{\sin \pi (S_3+T_3)}\rb A^E_4(-G_3-1,M_{23}) A^E_4(S_2,M_{23}-G_2-G_3-1) \times\nn\\
& &\times _3F_2(-S_3,-G_3,-G_2-S_3-T_3-1;-S_3-T_3,-G_3-T_2 ;1)\nn\\
&+& \lb 1- \frac{\sin \pi (M_{23}+S_3+T_3)}{\sin \pi (S_3+T_3)}\rb A^E_4(-G_3-1,M_{23}) A_4(T_2,M_{23}-G_2-G_3-1) \times\nn\\
\label{eq10loong}
& &\times _3F_2(-G_3,M_{23}-G_2-G_3,-T_3;-S_3-T_3,-G_3-S_2;1).
\end{eqnarray}
\end{widetext}
In Eq. \eqref{eq10loong} above all the $A^E_4$ amplitudes entering the expression on the right-hand side are Euler amplitudes, given by the expression~\eqref{eq2}, with $\Gamma$ the usual Euler Gamma function. Equation \eqref{eq10loong} is fully permutation-invariant in the momenta.

The expressions of the $A_5(k_i)$ amplitude in terms of the hypergeometric $_6F_5$ or $_3F_2$ functions analytically continue the function outside the region of convergence of the Koba-Nielsen integrals. Nonetheless, because these expressions for the amplitude involve hypergeometric functions at the critical point $z=1$, they will not always be defined for all values of the kinematic parameters $S_{2,3}$ $T_{2,3}$, $M_{23}$ in the complex plane. However, using the integration techniques above it is possible to derive alternative expressions for the amplitude in terms of hypergeometric functions that will be defined for a given regime of the parameters. Thus, the $A_5(k_i)$ amplitude has many closed-form presentations in terms of hypergeometric functions. We should emphasize that these expressions will always agree at values of the kinematic parameters where two or more presentations are defined.

Eq. \eqref{eq10loong} above of course matches the expressions for the five-point amplitude that can be derived in other ways, e.g. by computing momentum-ordered disk amplitudes \cite{Kitazawa1987Effective,Barreiro:2005hv,FairlieJones}. Note however that the 12 terms in Eq. \eqref{eq10loong} do not each correspond to one of the 12 momenta orderings, rather the presentation with each term corresponding to one of the momenta orderings is a different presentation of the amplitude, as explained above. Also note that changing presentations can change the number of terms, for e.g. the alternate expression for $A_5(k_i)$ we will give below has 14 terms.

Let's now give an example of an alternate presentation of the scattering amplitude, which can be useful for computing the small $\alpha'$ limit of the scattering amplitude. This presentation is obtained by separately computing some of the integrals on the subdomains in Eq. \eqref{eqR2}, rather than relating them by coordinate transforms to the integrals~\eqref{FTtrick11},~\eqref{FTtrick22}. Directly computing the integrals on the $(-\infty,0)\times(1,\infty)$ and $(1,\infty) \times (-\infty,0)$ intervals gives a contribution of
\begin{widetext}
\begin{eqnarray}
& &-\frac{2 \sin (\pi M_{23}) \sin (\pi T_2)  A_4^E(M_{23},S_3+T_3)  A_4^E(T_2,G_3+S_2)}{\cos \pi (S_2+T_2)-\cos \pi (2G_3+S_2+T_2)} \, _3F_2(-G_3,M_{23}-G_2-G_3,-T_3;-S_3-T_3,-G_3-S_2 ;1)\nn\\
& &-\frac{\sin (\pi S_3) \sin (\pi T_2)}{\sin (\pi (S_3+T_3)) \sin (\pi G_2) }   A_4^E(S_3,T_3) A_4^E(T_2,M_{23}+S_2) \,_3F_2(-M_{23},S_3+1,-G_2;-M_{23}-S_2,S_3+T_3+2;1)\nn\\
& &+\frac{\sin \pi M_{23}}{\sin \pi (S_3+T_3)} \lb-\frac{\sin \pi T_3}{ \sin \pi (G_3+S_2)}+\frac{\sin \pi S_3}{\sin \pi (M_{23}+S_2)}\rb A_4^E(M_{23}+S_2+S_3+1,T_3) \times\nn\\
\label{eqaddthis}
& &\times A_4^E(S_2,M_{23}) \, _3F_2(S_2+1,M_{23}+S_2+S_3+2,-T_2;M_{23}+S_2+2,G_3+S_2+2;1),
\end{eqnarray}
\end{widetext}
plus the symmetrized expression under $2\leftrightarrow 3$. To derive the alternate expression for the amplitude we must add these two expressions to Eq. \eqref{eq10loong}, and subtract the previous contribution of the intervals, that is
\begin{eqnarray}
\label{eq17}
& &A_4^{(E)}(-G_2-1,S_2) A_4^{(E)}(T_3,-G_3 -1) \times\\
& &\times \,_3F_2(-M_{23},S_2+1,T_3+1;-M_{23}-S_3,-M_{23}-T_2;1)  \nn
\end{eqnarray}
and the $2\leftrightarrow3$ symmetrized contribution. The resulting expression for the amplitude is given in the ancillary Mathematica notebook, although it can also be easily worked out from Eqs. \eqref{eq10loong}, \eqref{eqaddthis} and \eqref{eq17} above. This expression is valid in the small $\alpha'$ limit in kinematic variables $S_{2,3}$, $T_{2,3}$, and $M_{23}$.

\emph{Small $\alpha'$ limit.}---  To show the applicability of the exact formulas derived above, we now take a certain kinematic limit, so that inner products $\alpha' k_1\cdot k_{2,3}$, $\alpha'k_4 \cdot k_{2,3}$ and $\alpha' k_2 \cdot k_{3}$ are small relative to 1. This can be achieved by sending $\alpha'\to 0$ while keeping the inner products $k_1\cdot k_{2,3}$, $k_4 \cdot k_{2,3}$ and $k_2 \cdot k_{3}$ fixed, or alternatively orienting the momenta so that the inner products go to zero with $\alpha'$ fixed.\footnote{The Cauchy-Schwarz inequality does not hold for spacelike vectors, so this is a valid limit for tachyon momenta.}  Note that because of momentum conservation $\sum_{i=1}^5 k_i=0$ and the fact that the tachyon mass squared is proportional to $1/\alpha'$, this kinematic limit necessarily implies that some of the other momenta inner products (such as e.g. $\alpha' k_3k_5$) will be order $\OO(1)$. Thus the momenta are no longer treated on equal footing in this limit, and indeed the resulting amplitude expression (Eq. \eqref{thisiseeq18} below) will no longer be symmetric under all $S_5$ momentum permutations, although expression Eq. \eqref{eq10loong} and the other alternative presentations are fully permutation symmetric. Let's introduce the notation $S_{2,3}\eqqcolon\alpha's_{2,3}$, $T_{2,3}\eqqcolon\alpha't_{2,3}$, $M_{23}\eqqcolon\alpha'm_{23}$, so that we take the small $\alpha'$ limit while keeping $s_{2,3}$, $t_{2,3}$, and $m_{23}$ fixed. Then, in this limit, the amplitude becomes
\begin{widetext}
\begin{eqnarray}
\label{thisiseeq18}
& &A^{(G)}_5(k_i) = -\frac{\pi^2 m_{23} (s_2  t_3+ s_3  t_2)}{2 (m_{23}+s_2+ s_3+ t_2+ t_3)} \alpha'^2 \\
& &+\frac{\pi^2 m_{23} \lsb t_3 ( 3s_2m_{23}+3s_2^2+ s_2 s_3+ s_2t_2+ s_3t_2)+ s_3  t_2 (3 m_{23}+s_2+3s_3+ 3t_2)+3 s_2  t_3^2\rsb}{4 (m_{23}+s_2+ s_3+ t_2+ t_3)} \alpha'^3 + \OO(\alpha'^4). \nn
\end{eqnarray}
\end{widetext}
This expression can be derived by small-$\alpha'$ expanding the $_3F_2$ hypergeometric functions in the alternate presentation of the amplitude (given in the ancillary Mathematica notebook), that follows from Eqs. \eqref{eq10loong}, \eqref{eqaddthis} and \eqref{eq17}. The details of this computation will be reported in~\cite{paper2}.

\emph{Conclusion.}--- We expect our analysis to generalize to $n$-point higher derivative corrections.  We note that this corresponds to a recursion for the types of color-dual scalar higher-derivative corrections that can be used to lift QFT Yang-Mills and Einstein-Hilbert gravity order by order in $\alpha'$ via $Z$-theory~\cite{Carrasco:2016ldy,Mafra:2016mcc,Carrasco:2016ygv}, and thus should in principle be relatable to other off-shell recursion, \`a la Berends-Giele~\cite{Mafra:2016ltu,Mafra:2016mcc}. The former demonstrates compatibility with color-dual higher derivative operators and the latter points to possible generalizations beyond products of scalar amplitudes.  We leave exploration of both these exciting issues to future work.

\emph{Acknowledgments.}--- B.S. would like to thank Jonathan Heckman, An Huang and Xiao Zhong for useful discussions, and John Joseph M. Carrasco for insightful discussions, support and encouragement. B.S. would like to thank Carlos Mafra and Oliver Schlotterer for useful comments on an early draft of this paper. B.S. is indebted to Djordje Radicevic, Matt Headrick, and the other participants at the virtual Brandeis Meetings, where an in-progress version of this work was presented. This work was supported by the Department of Energy under Award Number DE-SC0021485.

\end{document}